\documentclass[twocolumn,showpacs,prc]{revtex4}

\usepackage{graphicx}
\usepackage{dcolumn}
\usepackage{bm}

\begin{document}
\title{Biological Electric Fields and Rate Equations for  Biophotons}
\author{M. Alvermann and Y. N. Srivastava}
\affiliation{Department of Physics \& INFN, University of Perugia, Perugia, IT}
\author{J. Swain and A. Widom}
\affiliation{Physics Department, Northeastern University, Boston MA, USA}

\begin{abstract}

Ultraweak bioluminescence - the emission of biophotons - remains an experimentally
well-established, but theoretically poorly understood phenomenon. This paper presents
several related investigations into the physical process of both spontaneous biophoton
emission and delayed luminescence. Since light
intensities depend upon the modulus squared of their corresponding electric fields we first make  
some general estimates about the inherent electric fields within various biological systems. 
Since photon emission from living matter following an initial excitation (``delayed luminescence'') typically 
does not follow a simple exponential decay law after excitation we discuss such non-exponential decays
from a general theoretical perspective and argue that they are often to be expected and why.
We then discuss the dynamics behind some nonlinear rate equations, connecting them both
to biological growth rates and biophoton emission rates, noting a possible connection with
cancer. We then return to non-exponential decay laws seen for delayed luminescence in
an experimental context  and again note a possible connection with cancer.

\end{abstract}


\maketitle

\section{Introduction \label{intro}}
Emission of biophotons from living cells and tissues of plant and animal origin are by now very 
well established (see, for example \cite{VanWijk2001, Popp1988, PoppExp:1988, Anshu}). Emission of biophotons 
can be technically considered as a type of bioluminescence, although the observed emission 
of biophotons from biological tissues is much weaker than what is observed in normal
bioluminescence and also different than thermal radiation emitted by tissues at their respective normal
temperatures. There are two forms: a spontaneous one which occurs continuously and light
emission following an initial optical excitation, often referred to as ``delayed luminescence''  (DL) which
typically does not follow an exponential decay law. n this connection, we mention in passing that 
detailed studies of the associated (Poisson \& sub-Poisson) statistics that arise from non-linear 
rate equations giving rise to them have also been made \cite{ChangSqueezed:2002, Yan:2002, ChangJiin2008}.

In Sec.(\ref{pef}), we discuss the biological cell and the
nucleus within as cavities for resonating electromagnetic 
fields with the aim of investigation the idea that biophotons might initially reside 
in these cavities. Also, the frequencies and the magnitude of the mean electric
fields are estimated.
In Sec.(\ref{discussion}), we discuss theoretical reasons for
deviations from a linear rate equation and the resulting exponential decay law for fluorescence,
which does not fit well with measured DL intensity as a function of time.
In Sec.(\ref{Dynamics}), motivated by the logistic equation, 
we discuss non-linear rate equations and the dynamics generating them
for photon emission by nonliving systems and make a connection
to growth rates for size and mass of cells {\it etc.} In Sec.(\ref{gbm}), the
growth of a biological mass is analyzed and in Sec.(\ref{pccg})  connected to 
biophoton intensity and its
correlation with the cancer cell growth is presented via some experimental results. In Sec.(\ref{ld}),
hyperbolic decays are considered and again a connection with cancer is noted.
We close the paper with some concluding remarks in Sec.(\ref{con}).

\section{Electric fields and frequencies in Bio-systems\label{pef}}

Weak emission of biophotons (also known as ultraweak bioluminescence)
from living cells and tissues of plant and animal origin is by now very 
well established \cite{VanWijk2001, Popp1988, PoppExp:1988, Anshu}
and occur not only in the ultraviolet, but also the visible region of the
spectrum, but it's precise physical origins remain unclear and are
the subject of ongoing investigation. There is an enormous literature,
but recent reviews can be found in
the monographs \cite{popp2003,shen2006,chang2010}.
With no obvious excited molecules or atoms having been identified
as being their sources, we estimate some of the characteristic
frequencies one expects for natural biological cavity resonators.

Let   \begin{math} L \end{math} denote the length scale of a cavity 
containing a fundamental frequency 
\begin{math} \omega \end{math} and a dielectric constant 
\begin{math} \varepsilon \end{math}. These are related by   
\begin{equation}
\omega =\left(\frac{\pi c}{L\sqrt{\varepsilon}}\right).
\label{pef1}
\end{equation}
For a typical biological cell and nucleus we have, respectively, that 
\begin{equation}
L_{\rm cell}\sim 10^{-3}{\rm cm} \ \ \ \ 
{\rm and}\ \ \ \ L_{\rm nucleus}\sim 10^{-4}{\rm cm}. 
\label{pef2}
\end{equation}
It is thereby expected that 
\begin{equation}
\omega_{\rm cell}\sim {\rm infrared} \ \ \ 
{\rm and}\ \ \ \omega_{\rm nucleus}\sim  {\rm optical}. 
\label{pef3}
\end{equation}
It is interesting to note here the size of the electric fields associated with 
one photon, 
\begin{eqnarray}
\left(\frac{\varepsilon E^2L^3}{4\pi }\right)=\hbar \omega = 
\left(\frac{\hbar \pi c}{L\sqrt{\varepsilon}}\right), 
\nonumber \\ 
E=\frac{2\pi }{L^2}\sqrt{\frac{\hbar c}{\varepsilon^{3/2}}}\ .
\label{pef4}
\end{eqnarray} 
A cavity photon located in the cell and nucleus, respectively, have 
associated electric fields 
\begin{eqnarray}
1\ {\rm Gauss}\equiv 299.792458\left(\frac{\rm volt}{\rm cm}\right),
\\ \nonumber
E^{\rm 1\ photon}_{\rm cell}\sim 3.5\times 10^{-2}\ {\rm Gauss}
\sim 1\ \left(\frac{\rm kilovolt}{\rm meter}\right),
\nonumber \\ 
E^{\rm 1\ photon}_{\rm nucleus}\sim 350 \ {\rm Gauss} 
\sim 0.1\ \left(\frac{\rm megavolt}{\rm meter}\right). 
\label{pef5}
\end{eqnarray} 
While the photon frequencies of the biological cavity modes in the cell and 
in the nucleus are in agreement with experiments, the estimates of the electric 
fields here presented are lower than previously reported \cite{Popp1988, Zhang:2000, RattemeyerDNA:1981}. 

\section{Discussion on non exponential decay laws \label{discussion}}

It is often thought that the luminescence following an initial excitation
should be exponential, but this invariably comes from some circular
reasoning: either ``this is well-known to be the case'' or ``this is what
one would expect from $dN/dt=-\lambda N$ where $N$ is the excited
population''. Deviations from exponential decay are actually quite
well-known \cite{Fonda}, though relatively absent from most textbooks.
Notable exceptions are the books by Ballantine \cite{Ballantine:1990}
and Merzbacher \cite{Merzbacher}. To quote from Merzbacher
({\em op. cit.} p513), ``...the fact remains that the exponential decay 
law, for which we have so much empirical support in radioactive
decay processes, is not a rigorous consequence of quantum mechanics
but the result of somewhat delicate approximations''. Among those approximations
is that the initial state be coupled to a large number of final states with
similar energies. For a treatment of systems decaying into small numbers
of final states and the attendant failure of the exponential ``law'', 
see, for example \cite{Dittes:2000}. Also of interest is \cite{Gaemers:1998}.

In fact, there are many simple physical systems which instead display hyperbolic
decay laws. Typical examples are those which involve the excitation of
pairs in the medium, which then recombine to emit light. This naturally
gives decay laws which one would expect classically to obey 
$dN/dt=-\lambda N^2$. Note that this is a purely classical result and
does not require coherent effects between the excited states, which would
also be expected to give the same decay law. For a review of condensed
matter analogs of hyperbolic delayed luminescence (DL) in living things,
see \cite{Scordino}. Interestingly, in systems like CdS,
the hyperbolic delayed luminescence depends strongly on the size of
the grains, and in the nematic liquid crystal {4-methoxybenzylidene-49-n-butylaniline} (MBBA) it is present in the crystalline form, but disappears on
melting \cite{Scordino}. In other words, the character
of delayed luminescence is not simply a matter of chemical composition, but
can depend strongly on the form the material takes. Interestingly, the trend 
seems to be towards higher DL in systems exhibiting higher degrees of
structure, a fact which seems relevant for biological systems.

Approximately hyperbolic decay laws also arise in correlated many-soliton
states \cite{Brizhik} which, again, may be relevant for biological systems.

It should be noted on general grounds, strict exponential decay is impossible
in quantum mechanics. Khalfin showed \cite{Khalfin} as far back as 1958 that
the Paley-Wiener
theorem, together with analyticity, forbids exponential decay at large times.

There is also a simple physical argument. Going to the energy representation from
the time representation one finds ({\em i.e.} taking a Fourier transform
of the amplitude, which would be proportional to $\exp{(-\frac{1}{2}\lambda t)}$)
a Lorentzian which immediately gives two problems: the tail goes
to infinity, so that a system with an initially finite energy could be found to
have arbitrarily large energy. Indeed, the Lorentzian (squared) does not have
a finite integral, so is unacceptable as a probability distribution for energy
on physical grounds.

There is a theorem \cite{zerotime} that if $P(t)$ is the probability
that a system with a finite mean energy remains excited -- that is, the survival probability -- must
satisfy $dP(t)/dt=0$ at $t=0$, a property shared by neither the exponential
nor hyperbolic decays, so they must at best be approximations where they
do seem to work. This can also be seen directly from the textbook ``derivation''
of exponential decay before any approximations are made.

Thus we see that a strictly exponential decay law fails at both large
and short times.

For the hyperbolic decay law, the same normalization problem in 
the energy representation is clearly present: the integral of $1/(t-a)$ from
zero to infinity is infinite and, again, we find that a strictly hyperbolic decay
law is unphysical, so nonlinear rate equations of this type must be
approximations. 

Weron and Weron \cite{Weron:1984} have argued
for a survival probability in general (for the cases where one might otherwise
derive, with approximations, an exponential decay) of the form $exp(-t^\alpha)$ where 
$\alpha>0$ and $0<\alpha<1$. It is even possible to have oscillations
in decay \cite{oscillations}.

Since one in general expects deviations from either exponential or
hyperbolic decays, any and all experimental data are welcome -- the form
of a decay law can, and indeed must, be more complicated than a simple
exponential or hyperbolic decay law, however well those models may fit
data over a restricted time interval.

In the following sections we shall discuss some non-linear rate equations 
that have been successfully employed for diverse biological systems.

\section{Dynamics behind some simple non-linear rate equations\label{Dynamics} }

Here we shall discuss some simple non-linear rate equations and the dynamical reasons behind
them. The linear rate equation where the rate is proportional to the number itself 
leads of course to an exponential growth or an exponential decay. But as in all practical systems, some non-linearity
is bound to be present giving rise to a non exponential in time behavior.

The best studied example is that of the laser. Here, if the mean photon number rate equation
were linear, the number of laser photons would increase exponentially. Of course, that cannot be otherwise
or we would need an infinite source of energy, hence there must be some dynamical mechanism to
saturate the number. The solution to this problem was first given by Willis E. Lamb. The famous Lamb
equation for light intensity $I(t)$ \cite{MW} may be written as 

\begin{equation}
\label{D1}
\frac{dI(t)}{dt} = +\nu [ a - I(t)] I(t). 
\end{equation} 
The second term on the right hand side of Eq.(\ref{D1}) arises dynamically through the creation 
and annihilation of two-photons at a time, just as the first term is related to the creation and annihilation 
of single photons. The parameter $a$ is called the pump parameter and its sign is crucial in determining
the steady state value of $I$. 

If $a\leq 0$, the steady state value of $I$ (determined by the vanishing of the left side of Eq.(\ref{D1}))
is $I_{SS}\to 0$. Physically, for negative pump parameters there is no laser activity. On the other hand though,
for $a>0$, $I_{SS} \to a$ and hence the laser intensity increases linearly with $a$.

Also, the innocent looking Eq.(\ref{D1}) has buried in it a (second order) phase transition wherein $a$ acts as 
the order parameter. This is easily seen by considering $I_{SS}$ as a function of $a$. $I_{SS}$ is continuous
at $a=0$ but its derivative is not.

A simple model for a plethora of physical processes such as mean photon number, intensity, 
mass growth, magnetization {\it etc.} is provided by analogs of the Eq.(\ref{D1}) where the parameters $\nu$ and $a$
have different physical significance and their signs do play a crucial role in determining the fate of that physical
system.   

As discussed previously in Sec.(\ref{pef}), the frequency of a mode in a 
biological cell is inversely proportional to the length in accordance with
Eq.(\ref{pef1}). If the cell geometry fluctuates via the length scale 
\begin{math} L \end{math}, then the frequency of the photon oscillator will 
be modulated. Because of such a modulation, the cell cavity will emit or 
absorb two photons at a time (as in the dynamical Casimir effect) thus leading to the above rate equation.

Similar (non-linear) rate equations must exist for any living system (such as for its size, cell number {\em etc.}) 
where their initial growth may be rapid but eventually cease, resulting in a limiting value (such as the maximum size). 
In the following section, we shall discuss a concrete case of biophoton emission from soybeans as well as 
the rate equation governing the growth in mass of said soybeans.   

\section{Growth of Biological Mass \label{gbm}}

\begin{figure}
\centering
\includegraphics[width=8cm]{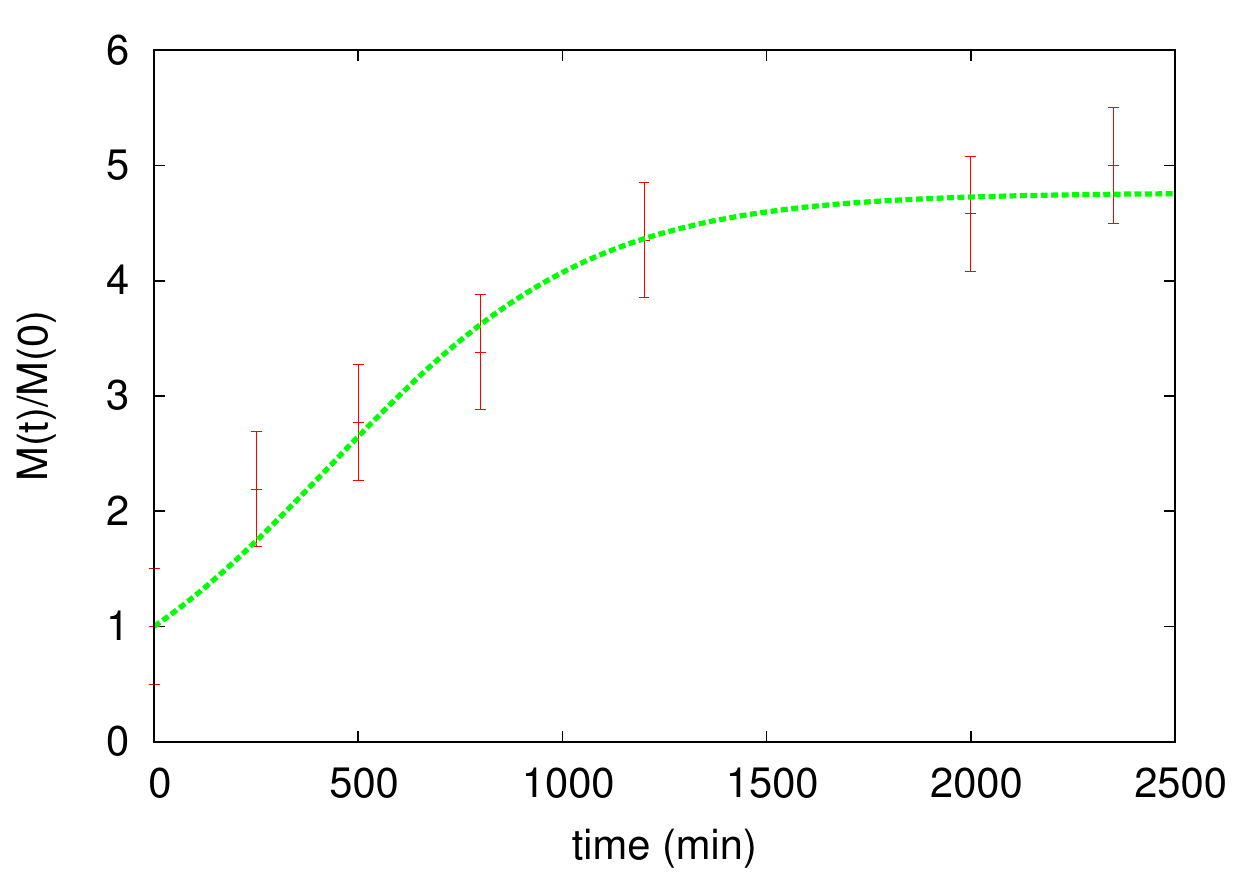}
\caption{
Data points exhibited from F.A. Popp {\it et al.}' \cite{Popp1988} showing the growing mass of soybeans as a 
function of time. The continuous curve is obtained using the logistic equation (Equation \ref{pbm2}) and parameters 
given in Equation \ref{pbm3}.}
\label{figure1}
\end{figure}

The growth of the biological soybean sprouts plus roots has been written as the 
solution of the logistic differential equation\cite{Verhulst:1845,Verhulst:1847}.
\begin{equation}
\frac{dM}{dt}=\nu \left( M-\frac{M^2}{M_\infty }  \right)
\label{pbm1}
\end{equation}
In this model, for low mass the growth rate is proportional to mass since the nutrients 
feeding the bean sprout is proportional to the mass. For higher mass 
the loss rate from soybean waste emission is proportional to the square of 
the mass because of second order reaction kinetics. 

With an initial mass \begin{math} M_0 \end{math}, the solution of Eq.(\ref{pbm1}) 
is given by 
\begin{equation}
M(t)=\frac{M_0 M_\infty}{M_0+(M_\infty -M_0)e^{-\nu t}}\ ,
\label{pbm2}
\end{equation}
where \begin{math} M_\infty \end{math} is the final steady state value of 
\begin{math} M(t\to \infty ) \end{math}.
In a measurement \cite{Popp1988} of the growth of bean sprout mass, 
one had 
\begin{eqnarray}
\nu =3.1\times 10^{-3} /{\rm min},
\nonumber \\ 
M_0=1.3\ {\rm gm},
\nonumber \\ 
M_\infty=6.19\ {\rm gm}.
\label{pbm3}
\end{eqnarray}
The comparison between the theoretical Eq.(\ref{pbm2}) and the experimental data on 
growing soybean sprouts is shown in FIG.\ref{figure1}. The agreement between theory and  experiment is satisfactory.

\section{Spontaneous Biophoton Emission and Cancer Cell Growth \label{pccg}}

\begin{figure}
\centering
\includegraphics[width=8cm]{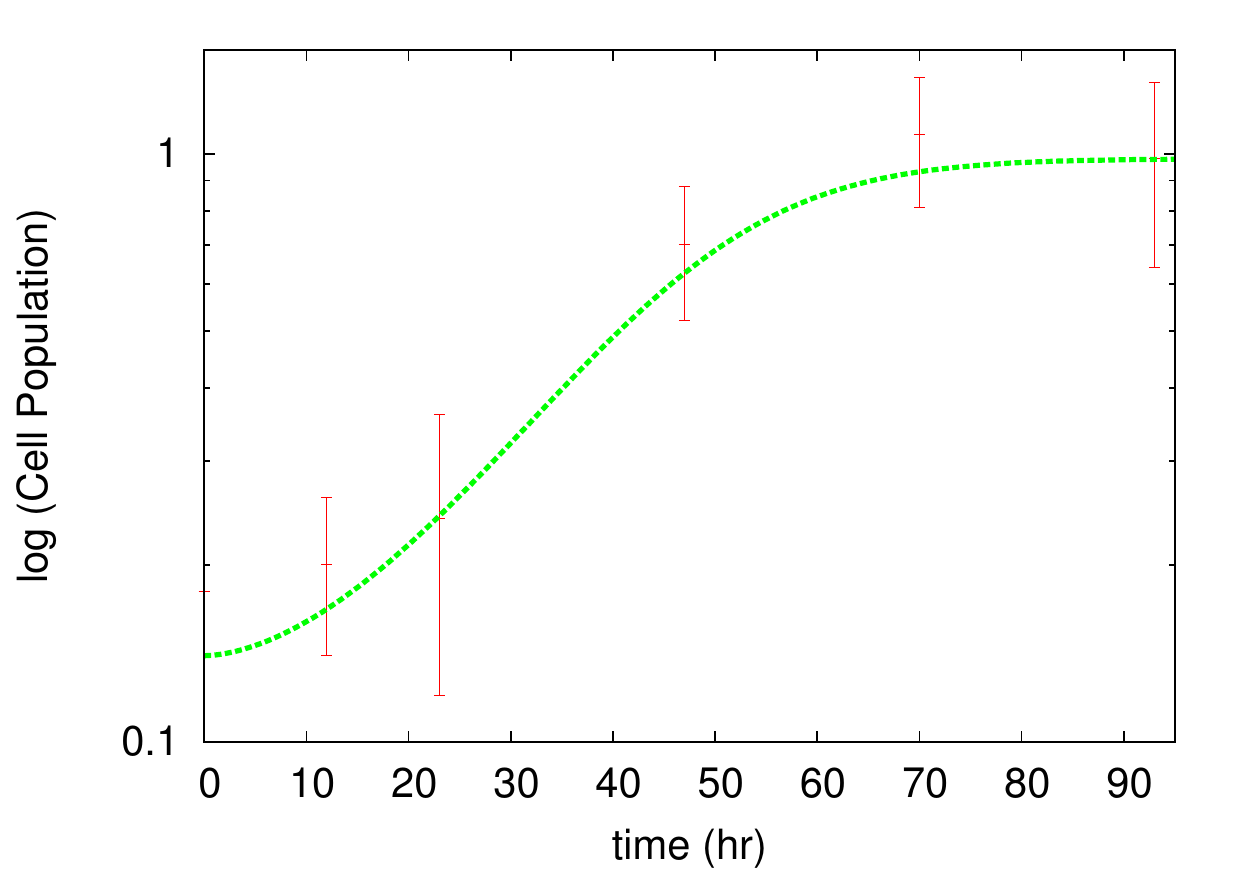}
\caption{Curve showing a stretched exponential curve
(Eqs.(\ref{pccg2}) and (\ref{pccg3})) with the experimental data of the growth of cancerous cell populations   
in arbitrary units as a function of time.}
\label{figure2}
\end{figure}

It is well-documented that at high cell densities, large differences in the spontaneous ultraweak photon emission 
exist between normal cells and tumor cells originating from the same parental tissue \cite{Schamhart, Takeda:1998}.
(In this section we confine our attention to the spontaneous biophoton emission and examine DL in section \ref{ld}.)
If the intrinsic rate of growth \begin{math} \nu \end{math} depends 
explicitly on time one finds a rate modulated logistic Eq.(\ref{pbm1}) of 
the form 
\begin{equation}
\frac{dM}{dt}=\nu (t) \left( M-\frac{M^2}{M_\infty }  \right).
\label{pccg1}
\end{equation}
The solution of Eq.(\ref{pccg1}) is 
\begin{eqnarray}
\eta (t)=\int_0^t \nu(s)ds,
\nonumber \\ 
M(t)=\frac{M_0 M_\infty}{M_0+(M_\infty -M_0)e^{-\eta (t)}}\ ,
\label{pccg2}
\end{eqnarray}
The growth curves of the esophageal cancer cell (TE9) mass have been carefully 
studied \cite{{Takeda:1998}}. We have fit the growth curves employing 
Eq.(\ref{pccg2}) with the stretched exponential form 
\begin{equation}
\eta (t)=(t/\tau)^r ,\ \ \tau \approx 6.31\ {\rm hr} 
\ \ {\rm and} \ \ r\approx 1.75.
\label{pccg3}
\end{equation}
The results are plotted in FIG.\ref{figure2}. The agreement with the 
stretched exponential form in Eqs.(\ref{pccg2}) and (\ref{pccg3}) is 
satisfactory. 

\begin{figure}
\centering
   \includegraphics[width=7cm]{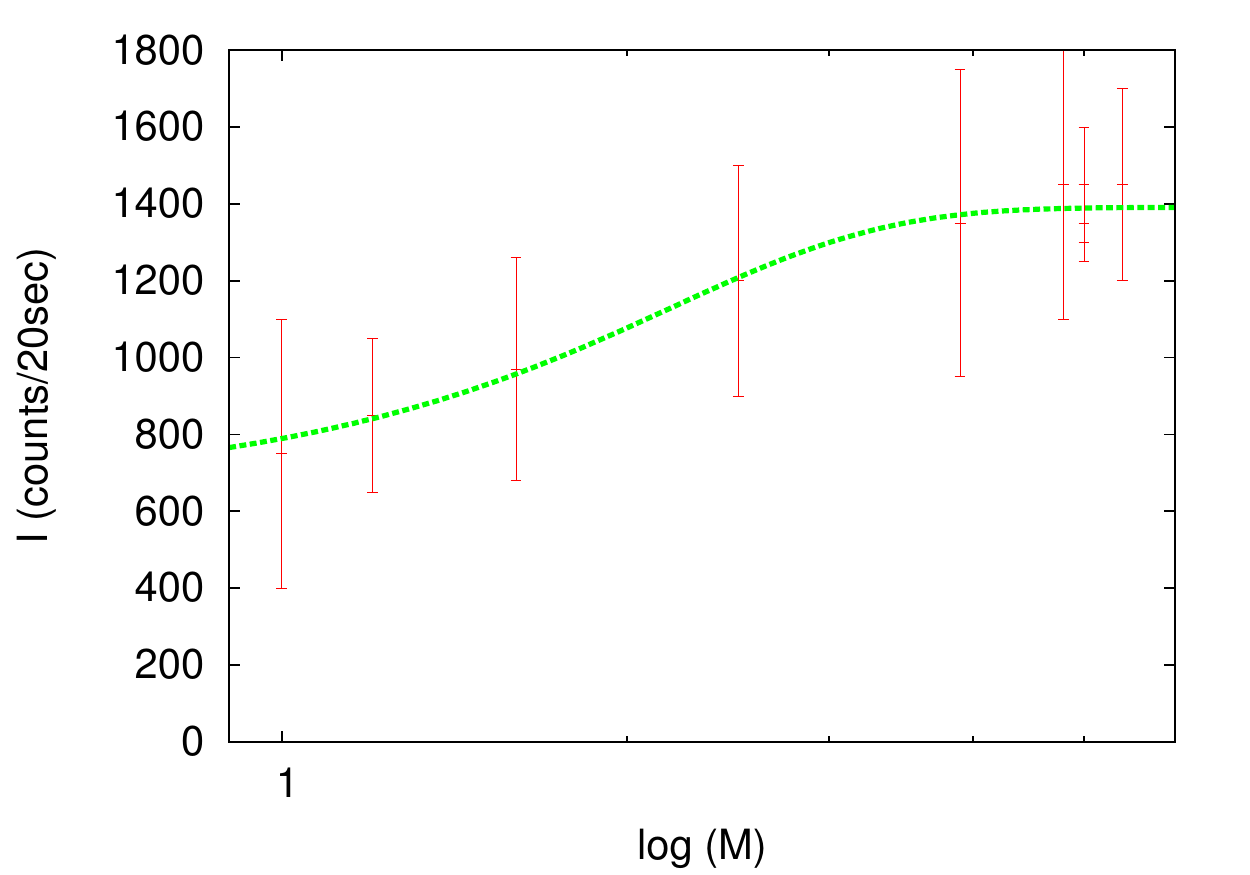}
      \caption{Curve showing the theoretical fit of Eqs.(\ref{pccg2}) 
and (\ref{pccg3}) with experimental data from Takeda {\it et al.} The  biophoton emission rate 
is shown as a function of the logarithm 
of the growing cancerous cell mass $M$ in arbitrary units.}
    \label{figure3}
\end{figure}

The biophoton emission intensity from the cancer cells as a function
of time has also been measured \cite{Takeda:1998}. One may thus obtain 
the biophoton intensity as a function of the number of cancer cells. 
A plot of the results is shown in FIG.\ref{figure3}. As the number of 
cancer cells increase, so does the biophoton emission rate of {\em each cell}
(see, for example, Fig.(19.11) in \cite{Popp1988}).

\section{Delayed Luminescence \label{ld}}

In this section we consider delayed luminescence and connect it with our 
earlier discussions.

A special case of Eq.(\ref{D1}) leads to a hyperbolic decay. Let us substitute
\begin{equation}
\label{LD1}
a = 0;\ \nu = \frac{\nu_o}{N(0)},
\end{equation}
so that Eq.(\ref{D1}) takes the form
\begin{equation}
\label{LD2}
\frac{dN(t)}{dt} = - (\frac{\nu_o}{N(0)}) N^2(t),
\end{equation}
whose solution is
\begin{equation}
N(t)=\frac{N(0)}{1+\nu_o t}
\label{LD3}
\end{equation}
A physical example \cite{Ruth:1981} of DL decay obeying the hyperbolic law is provided in Fig.(4).
\begin{figure}
\centering
\includegraphics[width=8.0cm, height=6.0cm]{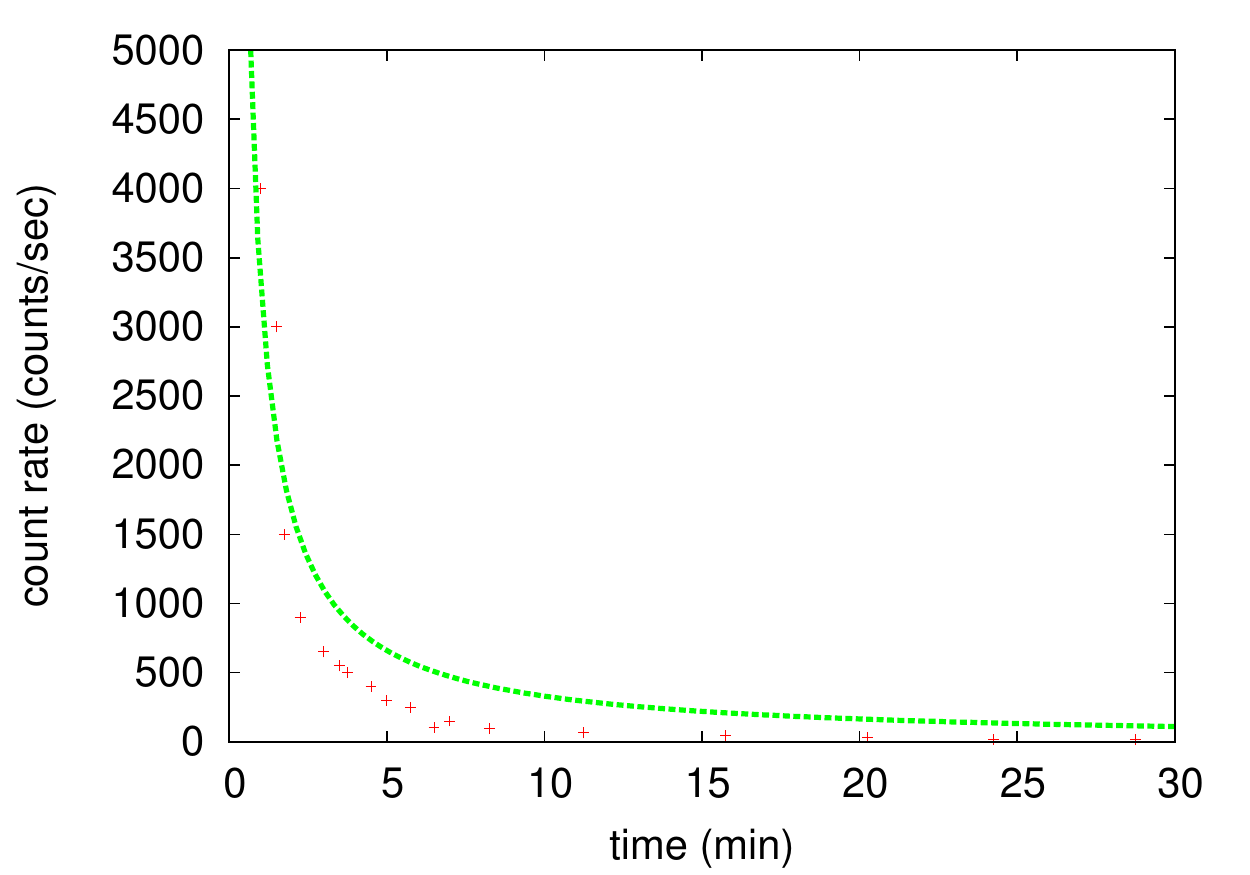}
\caption{Delayed luminescence intensity versus time for 
cells of {\em Bryophyllum daigremontanum}. The curve is the result of the fit of Eq.(\ref{LD3}) 
with the experimental data.}
\label{figure4}
\end{figure}

An extension of Eq.(\ref{LD3}) to a fractional power in time, in analogy to that of the 
stretched exponential as in the last section [see Eq.(\ref{pccg3})]  
\begin{equation}
\label{LD4}
M(t) = \frac{M(0)}{[1+\nu_o (t/t_o)^k]},
\end{equation}
has been made in \cite{Scholz:1988}, where light simulated biophoton reemission from
normal and cancerous cells are compared. They find a considerable difference in the value 
of the parameter $k$ for the two cases.

\section{Conclusions \label{con}}

An important element in the analysis of biophotons concerns the magnitudes of the electric
fields. Our estimates for the average electric fields in a cell and the nucleus - considered as a
cavity - show that these fields are quite large, see Sec.(\ref{pef}). For example, the field due to one such 
biophoton in a nucleus is only slightly smaller than the breakdown field in humid conditions.

In the rest of the paper we have considered various rate equations of particular relevance for 
biological systems and provided physical reasons behind those equations where ever possible,
see Sec.(\ref{Dynamics}), Sec.(\ref{gbm}) and Sec.(\ref{ld}). 
Theoretical arguments were presented to establish that the commonly employed exponential law 
for DL is not strictly tenable. A case of practical interest that is a delineation between the behavior of 
normal versus cancer cells is illustrated through differences in the values of parameters 
in their rate equations, see Sec.(\ref{pccg}) .

\end{document}